\documentclass[preprint]{elsarticle}
\usepackage{epsfig}
\usepackage{graphicx}
\usepackage{amssymb}
\newcommand{\AmS}{{\protect\the\textfont2
 A\kern-.1667em\lower.5ex\hbox{M}\kern-.125emS}}
\newcommand{\ba}{\begin{array}}
\newcommand{\ea}{\end{array}}

\def\beq{\begin{equation}}
\def\eeq{\end{equation}}

\def\bea{\begin{eqnarray}}
\def\eea{\end{eqnarray}}

\def\cl{{\rm c}}
\def\sg{{\rm s}}
\def\bg{{\rm b}}

\def\beq{\begin{equation}}
\def\eeq{\end{equation}}

\def\bea{\begin{eqnarray}}
\def\eea{\end{eqnarray}}

\begin{document}


\title{\Large 
A correlated-cluster  model
 and
  the ridge phenomenon\\ in hadron-hadron collisions
}




\author[ific]{Miguel-Angel Sanchis-Lozano}
\ead{Miguel.Angel.Sanchis@ific.uv.es}

\author[cern,uta]{Edward Sarkisyan-Grinbaum\corref{cor1}}
\ead{sedward@cern.ch}

\address[ific]{Instituto de F\'{\i}sica
Corpuscular (IFIC) and Departamento de F\'{\i}sica Te\'orica \\
\it Centro Mixto Universitat de Val\`encia-CSIC,
Dr. Moliner 50, E-46100 Burjassot, Valencia, Spain}

\address[cern]{Experimental Physics Department, CERN, 1211 Geneva 23, 
Switzerland}

\address[uta]{Department of Physics, The University of Texas at Arlington,
Arlington, TX
76019, USA}

\cortext[cor1]{Corresponding author}



\begin{abstract}
A study of the near-side ridge phenomenon in hadron-hadron
collisions based on a cluster picture of multiparticle production is
presented.  The near-side ridge effect is shown to have a natural
 explanation in this context provided that clusters are produced
 in a correlated manner
 in the collision transverse plane.
 \end{abstract}






\maketitle



\section{Introduction}

The study of multiparticle production in high energy hadron collisions
has proven to be a useful tool to explore the soft regime of 
the strong interaction dynamics \cite{book,DeWolf:1995pc}. 
In particular,
particle correlations are known to provide crucial information about 
the underlying mechanism of the multiparticle production process 
to be ultimately
interpreted in terms of QCD \cite{book,Dremin:2000ep}. 
 Moreover, the analysis of particle correlations has been
 shown 
 to reveal signals of non-conventional physics  
\cite{Sanchis-Lozano:2015eca,HS:IJMP}.

 In recent years big efforts have been devoted to the study of 
two-particle correlations in the search for collective phenomena 
(see \cite{CollEff-rev} for a review). Two-particle correlations 
 are analyzed in a two-dimensional azimuthal $\Delta \eta$-$\Delta \phi$ 
phase space, where $\Delta \phi$ and $\Delta \eta$ denote the difference 
of the 
azimuthal angle $\phi$ and the pseudorapidity $\eta$ of the two 
 selected
 particles, respectively \footnote{We consider a right-handed coordinate system with the 
$z$ axis along the beams' direction. Cylindrical coordinates are used in 
the transverse plane, $\phi$ being the azimuthal angle. The pseudorapidity 
is defined in terms of the polar angle $\theta$ as 
$\eta=-\ln{\tan{(\theta/2)}}$. The (longitudinal) rapidity is defined as 
$y=\frac{1}{2}\ln{(\frac{E+p_L}{E-p_L})}$ 
 and coincides with 
 the pseudorapidity for massless particles. Here, $p_L$ is the 
longitudinal (along the beam axis) component of the measured particle 
moment.}. Two-particle correlation function is defined as: 
 \begin{equation}\label{eq:corexp} C(\Delta \eta, \Delta \phi)= 
\frac{S(\Delta \eta,\Delta \phi)}{B(\Delta \eta, \Delta \phi)}\, , 
 \end{equation} 
where $S$ and $B$ denote particle pair 
distributions from the same event and from different events, 
representing the signal and background contributions, respectively \cite{Adare:2008ae}.
 
 Typically, a complex structure is observed
 for different energies and types 
 of
 colliding objects.
 On the one hand, 
 there is 
 a narrow peak centered at ($\Delta y \simeq 0$, 
$\Delta \phi \simeq 0$) 
 due to high transverse momentum 
($p_T$) clusters and jets, whereas a broader 
away-side 
Gaussian-type structure arises from
 the decay of lower $p_T$ clusters, resonances and 
 fragmentation, including Bose-Einstein correlations.
 Besides, an enhancement of two-particle
correlations is  also 
found at $\Delta \phi \simeq \pi$. Because of its 
extended shape 
as seen in the $\Delta \eta$-$\Delta \phi$ plot, it
is usually referred to as the {\em away-side ridge}. This effect
can be explained due to particle correlations coming from
momentum conservation in back-to-back jets.
 Another interesting structure is observed, namely 
 the long-range ($|\Delta \eta| \leq 5$) near-side
($\Delta \phi \simeq 0$) correlations yielding
a {\em near-side ridge}, whose study is the main objective
of this work.

 Long-range correlations are usually 
attributed to 
a collective 
hydrodynamical flow, and therefore the ridge structure is expected in 
nuclear collisions due 
to, e.g., an initial anisotropy that is imprinted 
on the azimuthal-angle distributions of final-state particles through the 
collective expansion of the medium \cite{CollEff-rev}. Another possible 
explanation is given 
by the color-glass condensate, where the two-gluon density is enhanced at 
small $\Delta \phi$, which still needs a collective flow boost to 
reproduce the observed ridge \cite{glasma-ridge}. Both mechanisms have, 
 however,  some 
 shortcomings 
 such as a locally thermalized medium,
which is required for the hydrodynamic 
flow  in order to 
 account for the near-side ridge phenomenon 
 \cite{hydro-review}. Finally, a third kind of explanation considers 
jet-medium interactions
where semihard partons can induce local fluctuations by energy loss 
in high density soft-parton fields, yielding
azimuthal asymmetries manifesting as a ridge structure \cite{Hwa:2010fj}.

 Unexpectedly, a long-range ridge structure has also been observed in
proton-nucleus \cite{ridge-pA,ridge-pp-pA-ATLAS} and, particularly, in 
proton-proton ($pp$) 
\cite{ridge-pp,ridge-pp-pA-ATLAS}  collisions, still requiring a 
definitive explanation.
The similarities between the correlations found in 
small systems and heavy-ion collisions suggest a common origin. 
Meanwhile, if hydrodynamics is successfully applied in the case of heavy ions,   
 the hydrodynamical explanation of the ridge effect in $pp$ 
collisions, 
even for high-multiplicity events where the effect has been 
actually observed,  
seems still
 unclear (however, see \cite{ppAAsimil}).
 This quite unexpected phenomenon still requires a careful study to 
establish its physical origin.

  Finally, it is worthwhile mentioning that 
the experimental analyses of azimuthal anisotropy usually involve
a Fourier series containing harmonics \cite{Voloshin:1994mz} 
up to fifth order, $\sum_{n=0}^5a_n\cos{(n\Delta \phi)}$, where the 
coefficients $a_n$ have to be 
interpreted and estimated by theoretical models.

In this Letter, we study two-particle correlations by invoking a 
simple 
two-step scenario for multiparticle production: the resulting multiplicity 
is given by the convolution of the distribution of particle emission 
sources (clusters/fireballs/semihard jets...) with the fragmentation/decay of 
 the sources. A correlated-cluster model (CCM) is 
 developed and
 applied throughout, 
 where 
 both 
  the 
 clusters and  
  the 
 final-state particles are 
  considered to be 
 emitted according to Gaussian distributions 
in 
rapidity and azimuth, encoding short-range and long-range correlations in 
both 
variables. 
 Our ultimate goal is to provide a common ({\it effective}) framework for 
both proton and heavy-ion collisions to deal with the near-side ridge 
phenomenon.
 Compact expressions are provided for $C(\Delta \eta, \Delta \phi)$. 

\section{Definitions and notation}

The general inclusive  two-particle correlation function is defined 
through the  Lorentz invariant
inclusive differential 
single and double cross sections $\sigma_{\rm in}^{-1}Ed^3\sigma/d^3p$ and 
$\sigma_{\rm in}^{-1}E_1E_2d^6\sigma/d^3p_1d^3p_2$,
respectively. Here,  $\sigma_{\rm in}$ denotes the 
inelastic cross section,
$E$ and $p$ denote the total energy and moment of particles, while
the subscripts 1 and 2 refer to the two considered particles. 
 As usual in this kind of analysis \cite{book},
 we will not 
distinguish between different species of particles, focusing
only on charged particles.

In terms of the rapidity ($y$) and 
 the azimuthal angle ($\phi$), the one-particle density  $\tilde{\rho}$ 
 and the two-particle 
 density 
$\tilde{\rho}_2$ are defined through
\begin{eqnarray}\label{eq:def2}
\tilde{\rho}(y,\vec{p}_T) &=& 
\frac{1}{\sigma_{\rm in}}\frac{d^3\sigma}{d^3p}= 
\frac{1}{\sigma_{\rm 
in}}\frac{d^3\sigma}{dyd^2p_T}=\frac{1}{2\sigma_{\rm 
in}}\frac{d^3\sigma}{dyd\phi 
dp_T^2} \, ,
\\ \nonumber
\tilde{\rho}_2(y_1,\vec{p}_{T 1},y_2,\vec{p}_{T 2}) &=& 
\frac{1}{\sigma_{\rm in}}\frac{d^6\sigma}{d^3p_1d^3p_2}= 
\frac{1}{\sigma_{\rm in}}\frac{d^6\sigma}{dy_1d^2p_{T 1}dy_2d^2p_{T 2}}=
\frac{1}{4\sigma_{\rm in}}\frac{d^6\sigma}{dy_1d\phi_1 dp_{T 
1}^2dy_2d\phi_2dp_{T 2}^2} \, , \nonumber
\end{eqnarray}
where $p_T$ ($p_T^2$) denotes the (square modulus of the) particle 
transverse momentum.

From the above expressions, let us 
define the normalized two-particle correlation function as
\begin{equation}\label{eq:C2}
C(1,2)=\frac{\tilde{\rho}_2(1,2)}{\tilde{\rho}(1)\tilde{\rho}(2)}\, ,
\label{eq:corfunction}
\end{equation}
 where the indices $1$ and $2$ stand for the set of kinematic variables 
 of the first and second particles of the pair,
 respectively.

On the other hand, it is customary 
integrating over $p_T^2$ on a suitable range denoted by $\Omega_T$
(determined by experimental cuts on events), and 
the single and two-particle densities become
\beq\label{eq:def2Omega}
\rho(y,\phi)=\int_{\Omega_T}dp_T^2\ \tilde{\rho}(y,\vec{p}_T)\ \ ,\ \ 
\rho_2(y_1,\phi_1,y_2,\phi_2)=\int_{\Omega_T}dp_{T 1}^2 dp_{T 2}^2\ 
\tilde{\rho}_2(y_1,\vec{p}_{T 1},y_2,\vec{p}_{T 2 }) \, .
\eeq

In order to match our theoretical approach to the 
 definition of 
 Eq.(\ref{eq:corexp}), 
 where the (pseudo)rapidity and azimuthal 
 differences are 
 involved, 
  we
 identify
 the 
 two-particle 
 distribution of the uncorrelated pairs by means of the two Dirac  
 $\delta$-functions, 
 i.e.
\begin{equation}
b(\Delta y, \Delta \phi)=\int dy_1dy_2d\phi_1d\phi_2\ 
\rho(y_1,\phi_1)\ \rho(y_2,\phi_2)\ \delta(\Delta y -y_1+y_2)\
\delta(\Delta \phi -\phi_1+\phi_2)  \, ,
\end{equation}
such that $\Delta y=y_1-y_2$ and $\Delta \phi=\phi_1-\phi_2$.

 In 
  its
 turn, the pair distribution of correlated pairs can be identified with
 \begin{equation}
s(\Delta y, \Delta \phi)=\int dy_1dy_2 d\phi_1d\phi_2\  
\rho_2(y_1,\phi_1,y_2,\phi_2)\ \delta(\Delta y -y_1+y_2)\
\delta(\Delta \phi -\phi_1+\phi_2) \, ,
\end{equation}
 where again 
 the Dirac $\delta$-functions 
 are
 incorporated. 
 
 Then, 
 the normalized 
 correlation function 
 is redefined 
 in the following way,
\begin{equation}\label{eq:c2bis}
C(\Delta y, \Delta \phi)= \frac{s(\Delta y, \Delta \phi)}{b(\Delta y, 
\Delta \phi)} \, ,
\end{equation}
 being  
 suitable for a 
 comparison with the experimental results obstained 
 using
 Eq.(\ref{eq:corexp}).

\section{Two-particle correlations in a cluster model}

It is usually accepted that particle production in soft hadronic interactions occurs
via an intermediate step of decaying ancestors/clusters/fireballs yielding 
final-state particles \cite{book,Dremin:2000ep}. 
It should be noted, however, that the 
 ``cluster'' 
 concept has to be understood
in a broad sense, i.e. a group of particles with some correlated 
properties. 

The Independent Cluster Model (clusters are produced in a non-correlated way) 
has been widely applied to the
study of hadron collisions 
(see \cite{Alver:2007wy, Alver:2008aa} and references therein).
 In 
 \cite{Chiu:2008ht}, the near-side ridge formation is studied 
in the framework of the so-called Correlated Emission Model for heavy ion collisions. In this
model, a semihard parton, emitted
in the primary collision, scatters when traversing the medium yielding a local energy flow. The correlation
between both (semihard parton and local flow) enhances the effect of soft emission, leading to the ridge formation. 
 On the other hand, our 
 approach,
 based on correlated cluster production,  
can be viewed as a rather model-independent approach to 
 ridge 
 formation, 
useful to deal with $pp$ collisions too.

 After integrating over the (square) cluster transverse momentum along the whole kinematically-allowed 
range, the single and two-cluster densities read 
 \beq\label{eq:rho2c}
\rho^{(\cl)}(y_\cl,\phi_\cl)=\int \ dp_{T \cl}^2\ 
\tilde{\rho}^{(\cl)}(y_\cl,\vec{p}_{T \cl})\ \ ,\ \ 
\rho_2^{(\cl)}(y_{\cl 1},\phi_{\cl 1},y_{\cl 2},\phi_{\cl 2})= 
\int dp_{T \cl 1}^2dp_{T \cl 2}^2\ \tilde{\rho}^{(\cl)}_2(y_{\cl 
1},\vec{p}_ 
{\cl 1},y_{\cl 2},\vec{p}_ {\cl 2}) \, .
\eeq

 Now we define particle densities depending on rapidity and azimuthal 
variables, coming from the decay of
a single cluster with rapidity $y_\cl$ and azimuthal angle $\phi_\cl$ 
leading to final-state particles: 
\begin{eqnarray}\label{eq:def2cl}
\rho^{(1)}(y,\phi;y_\cl,\phi_\cl) &=& \int_{\Omega_T}dp_T^2\ 
\tilde{\rho}^{(1)}(y,\vec{p}_T;y_\cl,\phi_\cl)  \, ,
\\
\rho_2^{(1)}(y_1,\phi_1,y_2,\phi_2;y_\cl,\phi_\cl) &=& 
\int_{\Omega_T}dp_{T 1}^2dp_{T 2}^2\ \tilde{\rho}_2^{(1)}(y_1,\vec{p}_{T 
1},y_2,\vec{p}_{T 2};y_\cl,\phi_\cl) \, ,
 \end{eqnarray}
 where  $\Omega_T$ again refers to the selected transverse momentum range 
 of final-state particles.

 Thus, the single particle density can be expressed as the convolution of the 
cluster density and the
particle density from a single cluster, i.e.
\beq\label{eq:rho}
\rho(y,\phi)=\int dy_\cl d\phi_\cl\ \rho^{(\cl)}(y_\cl,\phi_\cl)\ 
\rho^{(1)}(y,\phi;y_\cl,\phi_\cl) \, .
\eeq

Notice that all the kinematic variables appearing in the above expressions
(cluster and particle rapidities and azimuthal angles) are measured 
in the Laboratory Reference Frame (LRF) which coincides with the center-of-mass
frame of the hadron-hadron collision.

 To the extent that $\rho^{(1)}(y,\phi;y_\cl,\phi_\cl)$ is flat in 
the central rapidity region
and small transverse cluster momenta, we have $\rho(y,\phi)\ =\ \langle N_\cl\rangle\ \bar{\rho}^{(1)}$, 
where $\bar{\rho}^{(1)}$ stands for an average particle density for single cluster decay. Hence the
resulting single-particle density would be flat too and  its height
 to be  
 proportional to
the mean number of clusters per collision.

However, this approximation is quite rough as we are extending our study 
to the (pseudo)rapidities $| y | \lesssim 5$ and possibly to large 
cluster 
 transverse momenta.
 To this end, we introduce a function $E_1(y,\phi)$, which keeps the 
expected dependence on the rapidity and azimuthal variables of the emitted 
particles, so that 
the single-particle density becomes:
\beq\label{eq:rhoe}
\rho(y,\phi)\ =\ \langle N_\cl\rangle\ \bar{\rho}^{(1)}\ E_1(y,\phi) \, .
\eeq
  
 The $E_1(y,\phi)$ function is
normalized in such way that
\[
\int dy\ d\phi\ E_1(y,\phi)\ =\ 1 \, ,
\]
where integration is taken over the full available kinematic range of both 
the variables.

 Similarly, we introduce the product of the two single-particle 
distributions representing the mixed-event background, i.e.
\beq\label{eq:rhob}
\rho_{\rm mixed}(y_1,\phi_1,y_2,\phi_2)\ =\ 
\rho(y_1,\phi_1)\rho(y_2,\phi)\ =\ \langle N_\cl \rangle^2\ 
\bar{\rho}^{(1)2}E_1(y_1,\phi_1)E_1(y_2,\phi_2) \, ,
\eeq
which suggests to define
\beq\label{eq:E}
E_\bg (y_1,\phi_1,y_2,\phi_2)\ =\ E_1(y_1,\phi_1)E_1(y_2,\phi_2) \, .
\eeq

The two-particle density can also be written as
\beq\label{eq:rho2}
\rho_2(y_1,\phi_1,y_2,\phi_2)\ =\  
\int dy_\cl \phi_\cl\ \rho^{(\cl)}(y_\cl,\phi_\cl)\ 
\rho_2^{(1)}(y_1,\phi_1,y_2,\phi_2;y_\cl,\phi_\cl)\ +
\eeq
\[
\int dy_{\cl 1}dy_{\cl 2}d\phi_{\cl 1}d\phi_{\cl 2}\ 
\rho_2^{(\cl)}(y_{\cl 1},\phi_{\cl 1},y_{\cl 2},\phi_{\cl 2})\ 
\rho^{(1)}(y_1,\phi_1;y_{\cl 1},\phi_{\cl 1})\  
\rho^{(1)}(y_2,\phi_2;y_{\cl 2},\phi_{\cl 2}) \, .
\]
The first term on the r.h.s. corresponds to the emission 
of secondaries from a single cluster while the second term corresponds to
the emission of the two particles from two distinct clusters.

Therefore, we conclude for the two-particle density:
\beq\label{eq:corre}
\rho_2(y_1,\phi_1,y_2,\phi_2)\ =\  
\langle N_\cl \rangle\ \bar{\rho}^{(1)2}\ 
E_\sg^{\rm SR}(y_1,\phi_1,y_2,\phi_2)\ +\ \langle N_\cl(N_\cl-1) \rangle\ 
\bar{\rho}^{(1)2}E_\sg^{\rm LR}(y_1,\phi_1,y_2,\phi_2) \, , 
\end{equation}
where $E_\sg^{\rm SR}(y_1,\phi_1,y_2,\phi_2)$  stands for the 
short range (pseudo)rapidity correlations and 
$E_\sg^{\rm LR}(y_1,\phi_1,y_2,\phi_2)$ 
stands for the 
long range correlations  stemming from 
the two integrals of Eq. (\ref{eq:rho2}), respectively. 

 Needless to say, the above expression is mainly intended to describe the 
near-side effect using 
the CCM. Other kind
of  correlations (like particles inside jets or the away-side ridge) fall 
off the above description and
will not be considered in this Letter.

\subsection{Factorization hypothesis}
\label{factorhypo}

Taking into account that the rapidity $y$ and the azimuthal variable 
$\phi$ are orthogonal variables, we 
 tentatively assume that both $E_\bg (y_1,\phi_1,y_2,\phi_2)$ and 
$E_\sg(y_1,\phi_1,y_2,\phi_2)$ can be factorized as
\bea\label{eq:factor}
E_\bg (y_1,\phi_1,y_2,\phi_2) & = & E_\bg ^L(y_1,y_2) \cdot 
E_\bg ^T(\phi_1,\phi_2) \nonumber \, , \\
E_\sg(y_1,\phi_1,y_2,\phi_2) & = & E_\sg^L(y_1,y_2) \cdot 
E_\sg^T(\phi_1,\phi_2) \, ,
\eea
where the superscripts $L$ and $T$ denote the longitudinal 
and transverse parts, respectively.
 
Moreover, as usual in cluster models, we shall adopt Gaussian distributions  
in rapidity and azimuthal spaces for both cluster density 
and particle density from clusters 
\footnote{For a
isotropically decaying cluster with rapidity $y_c$, the single (massless) 
particle density  can be written as   $\rho^{(1)}(y;y_c)\ \sim\ 
\cosh{}^{-2}(y-y_c)$. As it is well known \cite{Bialas:1973kf}, it can be well 
approximated 
by a Gaussian of width $\delta_y \lesssim 0.9$.}, as developed below. 

On average, clusters should be isotropically
produced in the transverse 
plane of the primary hadronic collision
(even though anisotropy would be present in event-by-event fluctuations). 
Thus, only dependence on the rapidity variable remains in the single-cluster 
density,  
\beq\label{eq:cluster/part-rap}
\rho^{(\cl)}(y_\cl,\phi_\cl)\ \sim\ 
\exp{\biggl[-\frac{y_\cl^2}{2\delta_{\cl y}^2}\biggr]}\, ,
\eeq
where $\delta_{\cl y}$ denotes the rapidity correlation length 
for cluster production.
 On account of the plateau
structure of multiplicity distribution in pseudorapidity phase space, one 
may assume
that the dependence of $\rho^{(\cl)}(y_\cl,\phi_\cl)$ on $y_\cl$ is rather 
weak, i.e. $\delta_{\cl y}^2\gg 1$.
 Indeed, one can
empirically expect that 
$\delta_{\cl y}$ is basically determined by the rapidity plateau length of 
the
 single-particle distribution (for all charged particles).

\begin{figure}[t]
\begin{center}
\includegraphics[scale=0.33]{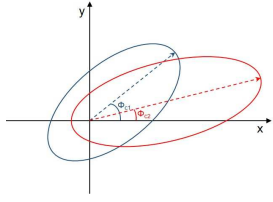}
\caption{Illustrative picture of two clusters produced in a primary 
hadron collision at the origin of the transverse plane 
with azimuthal angles $\phi_{c1}$ and $\phi_{c2}$,  
decaying into final state particles. 
Elliptic shapes are due to Lorentz 
boosts, although they could become somewhat distorted since they 
correspond
to the projection on the transverse plane
of boosted distributions in a three-dimensional space.}
\label{fig:drawing}
\end{center}
\end{figure}

 Now we turn to the single particle density 
$\rho^{(1)}(y,\phi;y_{\cl},\phi_{\cl })$. Since clusters are produced with 
some non-null (transverse) momentum, the initial isotropic distribution 
will be transformed into an elliptic shape depending on the cluster and 
emitted particle transverse velocities. (We note again that $\phi$ stands 
throughout for a variable in the LRF, which coincides with the center-of 
mass of the $pp$
 colliding system.)  
Hence a dependence on the cluster azimuthal angle $\phi_\cl$ should
remain in 
$\rho^{(1)}(y,\phi;y_\cl,\phi_\cl)$. 

As shown in the Appendix, such 
elliptic shape can be approximately expressed in terms of 
a Gaussian for highest boosted particles.
 Therefore 
 one gets
\beq\label{eq:part-rap-azim}
\rho^{(1)}(y,\phi;y_\cl,\phi_\cl)\ \sim\ 
\exp{\biggl[-\frac{(y-y_\cl)^2}{2\delta_y^2}\biggr]}\ 
\exp{\biggl[-\frac{(\phi-\phi_\cl)^2}{2\delta_{\phi}^2}\biggr]} \, .
\eeq 

 The parameter $\delta_y$, usually referred to as the cluster decay 
 ``width'', 
characterizes the (pseudo)rapidity separation of particles emitted in a 
single cluster decay; it has been
experimentally measured to be $\lesssim 1$ rapidity units \cite{Alver:2007wy}.
 Therefore, it turns out that $\delta_{\cl y}^2 \gg \delta_y^2$, in 
 accord with our previous discussion.

 For small azimuthal angles with respect to the 
cluster direction in the transverse plane  
\beq\label{eq:deltacpt}
\delta_{\phi}\ \sim\ \frac{1}{v_T\gamma_T}
\eeq
stands for the cluster decay 
 ``width'' regarding the azimuthal angle
instead of the rapidity variable. 
 Since the near-side ridge effect shows up for particles with transverse 
momentum of order of 1~GeV, we will take $\delta_{\phi} \simeq 0.14$ 
 radians 
 as a reference value (corresponding to a  pion with $p_T=1$~GeV emitted 
 at 
 rest in the cluster reference frame). 
 At low $p_T$, $\delta_{\phi}$ becomes large and the 
angular distribution remains almost flat
 (thereby the ridge disappears).
 Conversely, at higher $p_T$ values $\delta_{\phi}$  will
 decrease leading to a more pronounced
 peak at $\phi \simeq \phi_\cl$.
 Note also that one expects $\delta_{\cl \phi}^2 \gg \delta_{\phi}^2$ 
 since the particles from the boosted clusters 
 should be more collimated in azimuth than the clusters themselves (the 
 latter being quite more massive).

 On the other hand, we further assume that 
 the clusters are emitted in a correlated manner both in rapidity and 
azimuth.
 Thus,
 the two-cluster density 
is given by 
\beq\label{eq:nocorcluster}
\rho_2^{(\cl)}(y_{\cl 1},\phi_{\cl 1},y_{\cl 2},\phi_{\cl 2})\ \sim\ 
\exp{\biggl[-\frac{(y_{\cl 1}+y_{\cl 2})^2}{2\delta_{\cl y}^2}\biggr]} \ 
\exp{\biggl[-\frac{(\phi_{\cl 
1}-\phi_{\cl 2})^2}{2\delta_{\cl \phi}^2}\biggr]} \, ,
\eeq  
where $\delta_{\cl y}$ and $\delta_{\cl \phi}$ stand for the rapidity 
 and 
 azimuthal correlation lengths, respectively.
 Let us remark that Eq.~(\ref{eq:nocorcluster})
 can be regarded as a parameterization especially suitable to determine 
the near-ridge effect using the CCM. The physical origin of such azimuthal 
 correlations among clusters has to be provided by specific 
 models.

The underlying physical picture corresponds to cluster pairs 
emitted mostly with opposite rapidities 
but in the same hemisphere defined by 
the cluster velocity in the transverse plane. The correlation
strengths are determined by the $\delta_{\cl y}$ and
$\delta_{\cl \phi}$ parameters (not to be confused with
$\delta_y$ and $\delta_{\phi}$) respectively. 

The condition on the rapidity can be 
seen as a consequence of 
 longitudinal
 momentum conservation.
 Away-side particles can carry the remaining momentum due to a 
 non-exact 
 momentum balance by the two clusters along the beams' direction.
 The 
 azimuthal 
 condition 
 is 
 implemented in this version of the CCM 
 by hand\footnote{On the other hand, the Gaussian function 
$\exp{[-(\phi_{\cl 1}-\phi_{\cl 2}-\pi)^2/2\delta_{\cl \phi}^2]}$
would correspond to the back-to-back cluster emission in the transverse plane 
rather related to the 
the away-side ridge effect.} but should be attributed to a dynamical mechanism
developed in a concrete model. As 
 shown below, 
 the  requirement of azimuthal cluster correlations
  is definitely
needed in order to account for the near-side ridge 
effect according to the CCM.

In Fig.1 we illustrate the particle emission from two clusters
produced in the same primary hadron collision leading to different
elliptic shapes due to different Lorentz boosts. Clusters are assumed
to be correlated both in rapidity and azimuth according to 
Eq.(\ref{eq:nocorcluster}).

Now we integrate over rapidities and 
azimuthal angles using the Dirac's
$\delta$-functions. Lastly, we end up with a correlation function 
depending 
 on $e(\Delta y,\Delta \phi)$ yielding a 
 ``residual'' 
 dependence on 
 the rapidity and azimuthal variables:
\begin{equation}\label{eq:c2teo}
s(\Delta y,\Delta \phi)= 
\langle N_\cl \rangle\ \bar{\rho}^{(1)2}\ e_\sg(\Delta y,\Delta \phi) \, ,
\eeq
where $e_\sg(\Delta y,\Delta \phi)$ is defined as:
\beq\label{eq:g2}
e_\sg(\Delta y,\Delta \phi)\ =\ \int dy_1dy_2d\phi_1d\phi_2\ 
\delta(\Delta y -y_1+y_2)\
\delta(\Delta \phi -\phi_1+\phi_2)E_\sg(y_1,\phi_1,y_2,\phi_2) \, .
\eeq

Since one can reasonably expect that $\delta_{\cl y}^2 \gg \delta_y^2,\ 
 \delta_{\cl \phi}^2 \gg \delta_{\phi}^2$
 (as argued in Sect. \ref{factorhypo}),
 we write the short-range and long-range
pieces of the two-particle density as (see Appendix for $e$-functions):
 \begin{equation}\label{eq:sSR}
s^{\rm SR}(\Delta y,\Delta \phi)= \langle N_\cl \rangle\ \bar{\rho}^{(1)2}\ 
e_\sg^{\rm SR}(\Delta y, \Delta \phi) \, ,
\end{equation}
where 
\beq\label{eq:esSR}
e_\sg^{\rm SR}(\Delta y, \Delta \phi)\ \sim\ 
\exp{\biggl[-\frac{(\Delta y)^2}{4\delta_y^2}\biggr]}\ 
\exp{\biggl[-\frac{(\Delta \phi)^2}{4\delta_{\phi}^2}\biggr]} \, .
\eeq

On the other hand, 
\begin{equation}
s^{\rm LR}(\Delta y,\Delta \phi)= \langle N_\cl(N_\cl-1) \rangle\ \bar{\rho}^{(1)2}\ 
e_\sg^{\rm LR}(\Delta y, \Delta \phi) 
 \, ,
\end{equation}
where
\beq\label{eq:ebLR}
 e_\sg^{\rm LR}(\Delta y, \Delta \phi)\ \sim\  
 \exp{\biggl[-\frac{(\Delta \phi)^2}{2(2\delta_{\phi}^2+\delta_{\cl 
\phi}^2)}\biggr]} \, .
\eeq
Let us observe that for a Poissonian distribution of clusters 
$\langle N_\cl(N_\cl-1) \rangle\ =\langle N_\cl \rangle^2$.

Regarding the uncorrelated pairs, 
 one finds 
\beq\label{eq:b}
b(\Delta y,\Delta \phi)\ =\ \langle N_\cl \rangle^2\ \bar{\rho}^{(1)2}\ 
e_\bg(\Delta y,\Delta \phi) \, .
\eeq

Upon the integration over both rapidities, with $\Delta y$ fixed, one 
readily gets
\beq\label{eq:ee}
e_\bg(\Delta y,\Delta \phi)\ \sim\   
\exp{\biggl[-\frac{(\Delta y)^2}{4(\delta_y^2+\delta_{\cl y}^2)}\biggr]} 
\, . 
 \eeq
 Note that there is no azimuthal dependence in the above expression.


\section{Interpretation of the near-side ridge effect according to CCM}

In this section we show that the near-side effect emerges quite naturally 
from Eqs.(\ref{eq:sSR}) to (\ref{eq:ee}) according to the CCM. To this 
end, we write
\begin{equation}\label{eq:Cfinal}
C(\Delta y, \Delta \phi)=\frac{s^{\rm SR}(\Delta y,\Delta 
\phi)+s^{\rm LR}(\Delta y,\Delta \phi)}{b(\Delta y,\Delta \phi)}
=\ 1\ +\ \frac{h^{\rm SR}(\Delta y,\Delta \phi)}{\langle N_\cl \rangle}\ 
+\ \frac{\langle N_\cl(N_\cl-1) \rangle}{\langle N_\cl \rangle^2}\ 
h^{\rm LR}(\Delta \phi) \, ,
\end{equation}
where
\beq
h^{\rm SR}(\Delta y,\Delta \phi)\ =\    
\frac{e_\sg^{\rm SR}(\Delta y,\Delta \phi)}{e_\bg(\Delta y,\Delta \phi)}\ 
=\ 
\exp{\biggl[-\frac{\delta_{\cl 
y}^2}{4\delta_y^2(\delta_y^2+\delta_{\cl y}^2)}(\Delta y)^2\biggr]}\ 
\exp{\biggl[-\frac{(\Delta \phi)^2}{4\delta_{\phi}^2}\biggr]} \, ,
\eeq
and
\beq
h^{\rm LR}(\Delta y,\Delta \phi)\ =\    
\frac{e_{\sg}^{\rm LR}(\Delta y,\Delta \phi)}{e_{\bg}(\Delta y,\Delta 
\phi)}\ 
\simeq\  
  \exp{\biggl[\frac{(\Delta y)^2}{4(\delta_y^2
   +\delta_{\cl y}^2)} \biggr]}\ 
\exp{\biggl[-\frac{(\Delta 
\phi)^2}{2(2\delta_{\phi}^2+\delta_{\cl \phi}^2)}\biggr]} \, ,
\eeq
where the latter smoothly increases with the 
 (pseudo)rapidity separation $\Delta y$ 
 provided 
 that $\delta_{\cl y}^2 \gg 1$,
 as discussed above.

 For $\delta_{\cl y}^2 \gg \delta_y^2$ we find
\beq \label{eq:h2}
h^{\rm SR}(\Delta y,\Delta \phi)\ =\    
\exp{\biggl[-\frac{(\Delta y)^2}{4\delta_y^2}\biggr]}\ 
\exp{\biggl[-\frac{(\Delta \phi)^2}{4\delta_{\phi}^2}\biggr]}
\eeq
and
\beq\label{eq:h2bis}
h^{\rm LR}(\Delta y,\Delta \phi)\ \simeq\  
\exp{\biggl[-\frac{(\Delta 
\phi)^2}{2(2\delta_{\phi}^2+\delta_{\cl \phi}^2)}\biggr]}
\eeq
since there is an almost complete cancellation of the rapidity 
 dependence; only 
 the dependence on the azimuthal difference survives the 
 ratio
 leading to the rise of a ridge at small $\Delta\phi$.
 The physical reason is that $\delta_{\cl y}$ is provided by the plateau  
      length in the multiplicity pseudorapidity distribution (hence 
large), 
 while $\delta_{\cl \phi}$ is assumed to be quite smaller, as  
 shown 
 below.
   
 Indeed, the angular dependence of the second-order Fourier harmonic 
contribution ($\cos(2\Delta\phi)$) to the near-side correlation function 
can be approximated by the Gaussian of Eq.~(\ref{eq:h2bis}) for small 
$\Delta \phi$. 
 By Taylor-expanding the exponential 
 about $\Delta \phi \simeq 0$,
 and requiring 
 $\cos(2\,\Delta \phi)\simeq \exp\left[-(\Delta 
 \phi)^2/2(2\delta_{\phi}^2+\delta_{\cl \phi}^2)\right]$, 
 we get
 \beq
 2\delta_{\phi}^2+\delta_{\cl \phi}^2\ \simeq\ 0.25.
 \eeq

For high $p_T$ enough (therefore $\delta_{\phi}^2$ small) the above condition
leads to 
\[ \delta_{\cl \phi}\ \lesssim\ 0.5\ \  \mathrm{(radians)}, \]
in good agreement with findings from \cite{Chiu:2008ht,Chiu:2012rx}
and experimental measurements \cite{Agakishiev:2010ur}. Let us stress
that the long-range correlations in $h^{\rm LR}$ are
consequence of the azimuthal cluster correlations shown above, whereas
 the 
(pseudo)rapidity correlations play a minor role in the 
near-side ridge effect, as also pointed out in \cite{Chiu:2012rx}.

 Finally, 
 note that the weight of $h^{\rm LR}$ relative to the other terms
of Eq.(\ref{eq:Cfinal}) should 
increase as $\langle N_\cl \rangle$
increases. 
This can explain why the near-side ridge effect shows up at
larger multiplicity (hence larger $\langle N_\cl \rangle$).


\section{Summary}

We have studied the near-side ridge effect observed in nuclear and
 hadronic collisions at RHIC and LHC. 
 The study is carried out in the context of 
 the correlated cluster model (CCM) 
 using Gaussian distributions in azimuth and rapidity both for clusters 
 and final state hadrons, encoding short-range and long-range 
 correlations.

In order to reproduce the 
ridge phenomenon representing two-particle correlations 
at small azimuthal difference $\Delta \phi$  over a wide 
(pseudo)rapidity range, 
clusters have necessarilly to be emitted in a correlated 
 way in azimuthal space, but not in rapidity space.
 Let us stress that correlations among particles in single-cluster decays
 are not enough to account for the near-side ridge effect.
  Moreover, a relation for the second-order Fourier harmonic 
(related to the elliptic flow in a hydrodynamical scenario) is obtained
in agreement with experimental results.
Although the physical origin of
 cluster correlations could vary depending on the nature of the colliding 
bodies, the CCM provides 
a common framework to explain the ridge effect in proton-proton, 
proton-nucleus and heavy-ion collisions. 

\subsubsection*{Acknowledgements}

This work has been partially supported by MINECO under grant 
FPA2014-54459-P, and Generalitat Valenciana under grant 
PROMETEOII/2014/049.
 One of us (M.A.S.L.) acknowledges support from IFIC under grant 
SEV-2014-0398 of the ``Centro de Excelencia Severo Ochoa'' Programme.

\appendix

\section{Gaussian distributions and their convolutions}

\subsection{(Pseudo)rapidity dependence}

We will assume throughout that both clusters and particles stemming from 
clusters obey 
Gaussian distributions in rapidity space, i.e.
\beq\label{eq:cG}
\rho^{(\cl)}(y_\cl,\phi_\cl)\ \sim\ 
\exp{\biggl[-\frac{y_\cl^2}{2\delta_{\cl y}^2}\biggr]},\ \ \ 
\rho^{(1)}(y,\phi;y_\cl,\phi_\cl)\ \sim\ 
\exp{\biggl[-\frac{(y-y_\cl)^2}{2\delta_y^2}\biggr]} \, .
\eeq
respectively.

Upon integration over the cluster rapidity $y_\cl$, the $E_1^L(y)$ 
function, introduced
in Eq.(\ref{eq:rhoe}), reads
\beq\label{eq:E1}
E_1^L(y) \sim \int dy_\cl\ 
\exp{\biggl[-\frac{y_\cl^2}{2\delta_{\cl y}^2}\biggr]}\ 
\exp{\biggl[-\frac{(y-y_\cl)^2}{2\delta_y^2}\biggr]}\ \sim\ 
\exp{\biggl[-\frac{y^2}{2(\delta_y^2+\delta_{\cl y}^2)}\biggr]} \, .
\eeq

Hence, for two particles emitted from the two different clusters one gets 
for 
the longitudinal
part of the 
$E_\bg$ function, introduced in Eq.(\ref{eq:rhob}),
\beq\label{eq:E2}
E_\bg^L(y_1,y_2)\ =\ E_1^L(y_1) \cdot E_1^L(y_2)\ \sim\ 
\exp{\biggl[-\frac{(y_1^2+y_2^2)}{2(\delta_y^2+\delta_{\cl 
y}^2)}\biggr]}\, .
\eeq
Upon integration on both rapidities keeping the rapidity interval $\Delta 
y = y_1-y_2$ fixed, one gets
\beq\label{eq:2e}
e_\bg^L(\Delta y)\ \sim\ \exp{\biggl[-
\frac{(\Delta y)^2}{4(\delta_y^2+\delta_{\cl y}^2)}\biggr]}\, ,
\eeq

For two particles stemming from the same cluster with rapidity $y_\cl$  
\[
E_\sg^L(y_1,y_2) \sim \int dy_\cl\ 
\exp{\biggl[-\frac{y_\cl^2}{2\delta_{\cl y}^2}\biggr]}
\exp{\biggl[-\frac{(y_1-y_\cl)^2}{2\delta_y^2}\biggr]}
\ \exp{\biggl[-\frac{(y_2-y_\cl)^2}{2\delta_y^2}\biggr]}
\]
\beq
\sim\ 
\exp{\biggl[-\frac{\delta_{\cl 
y}^2(y_1-y_2)^2}{2\delta_y^2(\delta_y^2+2\delta_{\cl y}^2)}\biggr]}\ 
\exp{\biggl[-\frac{(y_1^2+y_2^2)}{2(\delta_y^2+2\delta_{\cl y}^2)}\biggr]}
\, .
\eeq

After integration using the Dirac's $\delta$-function, the above 
expression leads to  
\beq
e_\sg^{\rm SR}(y_1,y_2)\ \sim\ 
\exp{\biggl[-\frac{(\Delta y)^2}{4\delta_y^2}\biggr]} \, .
\eeq
Notice that $\delta_{\cl y}$ drops off in the last expression so that it 
can be referred to as a 
short-range correlation (SRC) contribution, as indicated in the superscript.

For two particles with rapidity $y_1$ and $y_2$
coming from two different (correlated) clusters with rapidities 
$y_{\cl 1}$ and $y_{\cl 2}$ respectively, we have (see 
Eq.(\ref{eq:nocorcluster}))
\[
E_\sg^L(y_1,y_2) \sim \int dy_{\cl 1}dy_{\cl 2} 
\exp{\biggl[-\frac{(y_{\cl 1}+y_{\cl 2})^2}{2\delta_{\cl y}^2}\biggr]} 
\exp{\biggl[-\frac{(y_1-y_{\cl 1})^2}{2\delta_y^2}\biggr]} 
\exp{\biggl[-\frac{(y_2-y_{\cl 2})^2}{2\delta_y^2}\biggr]}
\]
\beq
\sim\ 
\exp{\biggl[-\frac{(y_1+y_2)^2}{2(2\delta_y^2+\delta_{\cl y}^2)}\biggr]}\ .
\eeq
Using again the Dirac $\delta$-function, one gets
\beq
e_\sg^{\rm LR}(\Delta y)\ \sim\ {\rm const.}\,                                                     \, ,
\eeq
which corresponds to a long-range correlation (LRC), as  indicated in the 
superscript. 
 
In sum, we get two pieces with different behaviours (SRC versus LRC) 
in rapidity space:
\beq\label{eq:ebL}
e_\sg^{\rm SR}(\Delta y)\ \sim\ \exp{\biggl[-\frac{(\Delta y)^2}{4\delta_y^2}\biggr]}\ \ \ 
;\ \ \ e_\sg^{\rm LR}(\Delta y)\ \sim\  {\rm const.}
\eeq

 Had we used uncorrelated cluster production, i.e. $\rho_2^{(\cl)}\sim 
 \exp \left[ -(y^2_{\cl 1}+y^2_{\cl 2})/2\,\delta^2_{\cl y}\right]$, 
 our 
final results given by Eqs.~(\ref{eq:h2}) and (\ref{eq:h2bis}) would 
 remain the 
 same.

\subsection{Azimuthal dependence}

Clusters are supposed to be produced isotropically in the transverse plane of the 
hadron collision. Let $\phi_\cl$ be a particular value of the azimuthal 
variable for 
a cluster whose decay into particles is also assumed isotropic. 
Thus the azimuthal distribution $w(\phi^{\ast})$ in the cluster
rest frame should be a constant.
 
Under a Lorentz boost of velocity $v_T$, the angular distribution in the LRF
is given by \cite{Byckling:1971vca}
\beq\label{eq:polar}
w(\phi-\phi_\cl)=\frac{1}{\gamma_{T}[1-v_T^2\cos^2{(\phi-\phi_\cl)}]}\ 
f(\phi,g)\, .
\eeq 
We recall here that 
$\phi$ is measured in the LRF, i.e. the angular distribution  
is boosted by a Lorentz factor 
$\gamma_T=(1-v_T^2)^{-1/2}$ in the transverse plane. The $f(\phi,g)$ 
function stands for
\beq\label{eq:f}
f(\phi,g)\ =\ \frac{g\ \pm\ \sqrt{D}}{\pm \sqrt{D}} \, , 
\eeq
where $D=1+\gamma_T^2(1-g^2)\tan{}^2(\phi-\phi_\cl)$, with 
$g=v_T^{(\cl)}/v_T^{\ast}$ denoting the ratio of
the cluster transverse velocity $v_T^{(\cl)}$ in the LRF
and the particle transverse velocity $v_T^{\ast}$ measured in the cluster 
rest frame,  respectively.

For $g \geq 1$ all particles are emitted along the
forward hemisphere (defined by the cluster velocity) in the transverse plane
of the LRF.
In fact, this should be the case for particles, emitted with transverse 
momentum in the range $1 \leq p_T \leq 5$ GeV measured in the LRF. 
Therefore, we will 
consider that $g \approx 1$ and set 
$f(\phi,g)$ approximately  equal to a 
constant. Larger transverse momenta of particles would correspond rather 
to jet production instead of an intermediate-$p_T$ cluster production on 
 which we are focusing 
 in this work.
 
 In order to get simpler expressions at the end, 
 we approximate the azimuthal distribution by a Gaussian for small $\phi-\phi_c$ angles, 
namely, 
\beq 
w(\phi-\phi_\cl)\ \approx\ 
\exp{\biggl[-\frac{(\phi-\phi_\cl)^2}{2\delta_{\phi}^2}\biggr]},\ \ \ 
\delta_{\phi}\ \simeq\ \frac{1}{v_T\gamma_T} \, .
\eeq
It becomes apparent that large cluster transverse velocity  
leads to small $\delta_{\phi}$ and thereby
a more pronounced peak at $\phi \simeq \phi_\cl$, 
in accordance
with Eq.(\ref{eq:polar}). 

Admittedly, cluster emission boosted along the transverse 
plane has only been considered above, while 
the general situation should contemplate cluster and particle 
motions with velocity components
along the beam direction as well. However, the
main conclusion 
 of 
 the 
 above should remain valid.

In addition to the hypothesis of isotropically decaying clusters in their own rest frame,
we will assume axial symmetry for cluster production in the transverse plane, i.e.
\beq
E_{\bg}^T(\phi_1,\phi_2)\ \sim\ \mathrm{const.}\ \to\ e_{\bg}^T(\Delta 
\phi)\ \sim\ \mathrm{const.}\
\eeq

Thus, the distribution for two particles, emitted from the same cluster 
with azimuthal angle $\phi_\cl$ should obey 
\beq
\int d\phi_\cl\ 
\exp{\biggl[-\frac{(\phi_1-\phi_\cl)^2}{2\delta_{\phi}^2}\biggr]}\ 
\exp{\biggl[-\frac{(\phi_2-\phi_\cl)^2}{2\delta_{\phi}^2}\biggr]}\ 
\sim\ \exp{\biggl[-\frac{(\phi_1-\phi_2)^2}{4\delta_{\phi}^2}\biggr]}
\eeq
for small azimuthal angles. Therefore, regarding the azimuthal dependence we can write
\beq
e_\sg^{\rm SR}(\Delta \phi)\ \sim\ \exp{\biggl[-\frac{(\Delta 
\phi)^2}{4\delta_{\cl \phi}^2}\biggr]} \, .
\eeq

On the other hand, we will assume that 
clusters are produced in a correlated way according
to Eq.(\ref{eq:nocorcluster}). Hence 
for two particles with azimuthal angles $\phi_1$ and $\phi_2$
coming from two different clusters with azimuthal angles $\phi_{\cl 1}$ 
and $\phi_{\cl 2}$, 
 we will write
\beq
E_\sg^T(\phi_1,\phi_2) \sim \int d\phi_{\cl 1}d\phi_{\cl 2} 
\exp{\biggl[-\frac{(\phi_{\cl 
1}-\phi_{\cl 2})^2}{2\delta_{\cl \phi}^2}\biggr]} 
\exp{\biggl[-\frac{(\phi_1-\phi_{\cl 1})^2}{2\delta_{\phi}^2}\biggr]} 
\exp{\biggl[-\frac{(\phi_2-\phi_{\cl 2})^2}{2\delta_{\phi}^2}\biggr]}
\eeq
\[
\sim\ 
\exp{\biggl[-\frac{(\phi_1-\phi_2)^2}{2(2\delta_{\phi}^2+
\delta_{\cl \phi}^2)}\biggr]} \, ,
 \] 
that directly leads to
\beq
e_\sg^{\rm LR}(\Delta \phi)\ 
\sim\ \exp{\biggl[-\frac{(\Delta \phi)^2}{2(2\delta_{\phi}^2+\delta_{\cl 
\phi}^2)}\biggr]} \, ,
\eeq
which corresponds to a LRC, as indicated in the superscript.

\subsection{Final expressions}

 In sum, we find
  that the SRC and the LRC pieces of the $e_\sg(\Delta 
y,\Delta \phi)$ function
can be written as
\[
e_\sg^{\rm SR}(\Delta y, \Delta \phi)\ \sim\ \exp{\biggl[-\frac{(\Delta y)^2}{4\delta_y^2}\biggr]}\ 
\exp{\biggl[-\frac{(\Delta \phi)^2}{4\delta_{\phi}^2}\biggr]}
\]
and
\[
e_\sg^{\rm LR}(\Delta y, \Delta \phi)\  \sim\ 
\exp{\biggl[-\frac{(\Delta \phi)^2}{2(2\delta_{\phi}^2+\delta_{\cl 
\phi}^2)}\biggr]} \, .
\]

 Note that
$e_\bg(\Delta y, \Delta \phi)$ only retains dependence
on the rapidity variable for isotropic cluster production in the transverse 
plane,
 \[
e_\bg(\Delta y, \Delta \phi)\ \sim\ 
\exp{\biggl[-\frac{(\Delta y)^2}{4(\delta_y^2+\delta_{\cl y}^2)}\biggr]} \, 
.
\]

\end{document}